\documentclass[9pt,twocolumn,twoside]{osajnl}
\usepackage{graphicx}
\usepackage{epstopdf}
\usepackage{amsmath,amssymb,bbm,upgreek}
\usepackage{hyperref}

\usepackage{color}

\journal{ol} 

\setboolean{shortarticle}{false} 

\ifthenelse{\boolean{shortarticle}}{\colorlet{color2}{color2b}}{\colorlet{color2}{color2}} 

\title{13\,dB Squeezed Vacuum States at 1550\,nm from 12\,mW external pump power at 775\,nm}

\author[1]{Axel Sch{\"o}nbeck}
\author[2]{Fabian Thies}
\author[1]{Roman Schnabel}

\affil[1]{Institut f\"ur Laserphysik und Zentrum f\"ur Optische Quantentechnologien, Universit\"at Hamburg, Luruper Chaussee 149, 22761 Hamburg, Germany}
\affil[2]{Max-Planck-Institut f\"ur Gravitationsphysik (Albert-Einstein-Institut) and Institut f\"ur Gravitationsphysik, Leibniz Universit\"at Hannover, Callinstra{\ss}e 38, 30167 Hannover, Germany}

\affil[*]{Corresponding author: axel.schoenbeck@physik.uni-hamburg.de}

\dates{Compiled \today}

\ociscodes{}

\doi{\url{http://dx.doi.org/10.1364/ao.XX.XXXXXX}}

\begin{abstract}
Strongly squeezed light at telecommunication wavelengths is the necessary resource for one-sided device-independent quantum key distribution via fibre networks. Reducing the optical pump power that is required for its generation will advance this quantum technology towards efficient out-of-laboratory operation. Here, we investigate the second-harmonic pump power requirement for parametric generation of continuous-wave squeezed vacuum states at 1550\,nm in a state-of-the-art doubly-resonant standing-wave PPKTP cavity setup.  We use coarse adjustment of the Gouy phase via the cavity length together with temperature fine-tuning for simultaneously achieving double resonance and (quasi) phase matching, and observe a squeeze factor of 13\,dB at 1550\,nm from just 12\,mW external pump power at 775\,nm. We anticipate that optimizing the cavity coupler reflectivity will reduce the external pump power to 3\,mW, without reducing the squeeze factor. 
\end{abstract}

\setboolean{displaycopyright}{true}

\begin{document}

\maketitle
\thispagestyle{fancy}

\ifthenelse{\boolean{shortarticle}}{\ifthenelse{\boolean{singlecolumn}}{\abscontentformatted}{\abscontent}}{}

\section{Introduction}

Allocating suitable low-noise laser light of sufficient power in a well-defined spatial mode contributes most to the rather large footprint of current sources for strongly squeezed light. These sources use cavity-enhanced degenerate parametric down-conversion \cite{Yurke1984,Collett1985,Wu1986} in optically nonlinear crystals 
such as periodically poled potassium titanyl phosphate (PPKTP), 
for a review see also  \cite{Schnabel2017}. Reducing the required external pump power will allow for smaller laser devices, with smaller foot prints and less power consumptions and is a significant step towards a compactification of squeezed-light sources. Sources for strongly squeezed light form the basis of the, so far, only existing, experimentally proven concept of quantum key distribution (QKD) that does not need to trust the remote receiver. In such one-sided device independent QKD, any attacks on the receiver site, including flaws of receiver implementation and operation, are discovered during channel characterization, as demonstrated in \cite{Gehring2015}.\\
In \cite{Mehmet2011}, 12.3\,dB of squeezing at 1550\,nm was observed using a PPKTP squeezing cavity that was only resonant at 1550\,nm but not at the pump wavelength of 775\,nm, of which 171\,mW was required. 
It is well known that squeezing cavities that are not only resonant for the squeezed field but also for the pump field allow for lower external pump powers \cite{Wu1986}. 
In \cite{Stefszky2012}, a doubly-resonant traveling-wave squeezing resonator was realized and a squeeze factor of 11.6\,dB at audio sideband frequencies around 1064\,nm observed, using a pump power of 90\,mW. 
In \cite{Vahlbruch2016}, a doubly-resonant standing-wave squeezing resonator was realized and a squeeze factor of 15\,dB at MHz sideband frequencies around 1064\,nm observed, using a pump power of 16\,mW. 
The experimental challenge is to simultaneously achieve double resonance as well as (quasi) phase matching,
since the Gouy phase \cite{Lastzka2007} and dielectric mirror coatings (and a crystal length not being an integer multiple of the poling period) cause differential phase shifts.
One solution is the introduction of an adjustable dispersive component into the cavity, such as a rotatable anti-reflection (AR) coated window \cite{Stefszky2012} \cite{Pearl1999} \cite{McKenzie2006}  or adjustable gas pressure \cite{Wu1987}. Both examples, however, increase the complexity of the system, and, if the light needs to propagate through additional windows, additional optical loss is introduced, which reduces the achievable squeeze factor. 
Another possibility is to use a wedged crystal. Its translation perpendicular to the cavity mode changes the differential path length and allows for achieving double resonance and phase matching simultaneously \cite{Imeshev1998} \cite{Stefszky2010}. This approach, however, is not feasible if one crystal surface is used as a (spherical) cavity mirror for realising a compact 'half-monolithic' design.

Here, we report on the observation of a $13\,\mathrm{dB}$ squeezed vacuum state at $1550\,\mathrm{nm}$ from a doubly-resonant quasi-phase-matched standing-wave half-monolithic cavity that was pumped with just $12\pm1\,\mathrm{mW}$ of light at $775\,\mathrm{nm}$. The nonlinear medium inside the cavity was a $9.3\,\mathrm{mm}$ long PPKTP crystal. 
Double resonance and close to optimum quasi-phase-matching was achieved simultaneously without introducing an additional dispersive component. Coarse trial and error 
adjustment of the Gouy phase via the
cavity length together with fine tuning of the crystal temperature turned out to be sufficient.
Compensation of differential phase shifts from e.g.~mirror coatings is possible because the Gouy phase translates to different additional optical path lengths for the two fields \cite{Lastzka2007}. 

\section{Experimental setup}

\begin{figure}[htbp]
\centering
\vspace{-2mm}
\mbox{\includegraphics[width=\linewidth]{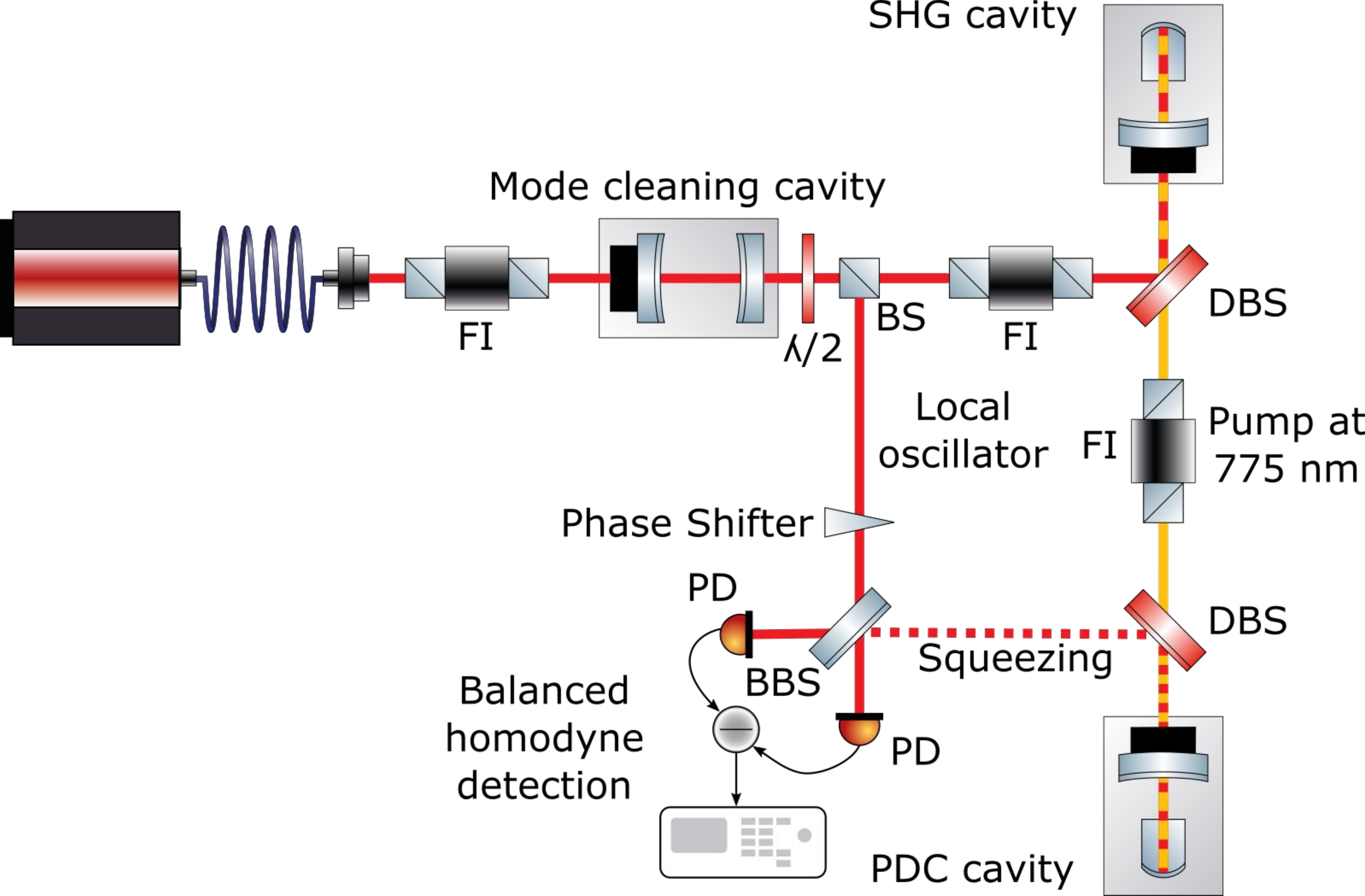}} 
\caption{Schematic of the experiment. The continuous-wave monochromatic laser light at 1550\,nm was filtered by a standing wave mode-cleaner cavity and partly up-converted to 775\,nm with a second harmonic generation (SHG) cavity. The latter field pumped cavity-enhanced type\,I degenerate spontaneous parametric down-conversion (PDC). Thus, a squeezed vacuum state at sideband frequencies around 1550\,nm was generated. A dichroic beam splitter (DBS) deflected the squeezed field towards a balanced homodyne detector (BHD). (B)BS: (balanced) beam splitter; PD: photo diode; FI: Faraday isolator.}
\label{Exp}
\end{figure}

Figure \ref{Exp} shows the schematic of the experiment. Continuous-wave light at $1550\,\mathrm{nm}$, which was provided by a fibre laser, was first  filtered by a standing-wave cavity and partially up-converted to 775\,nm by cavity-enhanced second-harmonic generation (SHG), using the design described in \cite{Baune2015}. The field at $775\,\mathrm{nm}$ was separated from the remaining field at $1550\,\mathrm{nm}$ with a dichroic beamsplitter (DBS) and used to pump degenerate spontaneous parametric down conversion (PDC) for the generation of squeezed vacuum states at 1550\,nm. The PDC process was enhanced by a half-monolithic standing-wave cavity consisting of a PPKTP crystal of 9.3\,mm length and a coupling mirror that was mounted on a piezo-electric element (PZT) \cite{Baune2015}. The coupling mirror had a reflectivity of 97.5\% at 775\,nm and 85\% at 1550nm. The back surface of the crystal had a high-reflection coating for both wavelengths. 
The length of the PDC cavity was actively stabilized by actuating on the PZT using the Pound-Drever-Hall locking method. The error signal was derived from the pump field. For this purpose, a phase modulation at $17.5\,\mathrm{MHz}$ was imprinted on the fundamental field at 1550\,nm and up-converted in the SHG. A DBS spatially separated the $775\,\mathrm{nm}$ pump field from the squeezed field and guided the latter to a balanced homodyne detector.\\
For balanced homodyne detection, the squeezed field was overlapped with the local oscillator on a balanced beam splitter (BBS). A photo diode (PD) was placed in each output port of the BBS. The photocurrents from the two PDs were subtracted from each other and the resulting voltage was fed to a spectrum analyser. The quadrature that is read out by the homodyne detector is defined by the input phase difference. For that reason, a phase shifter, consisting of a mirror that is mounted on a PZT, was placed in the local oscillator path. Applying a voltage to the PZT allowed us to detect the squeezed or anti-squeezed quadrature, respectively.

\section{Experimental results}

\begin{figure}[htbp]
\centering \vspace{-2mm}
\mbox{\includegraphics[width=\linewidth]{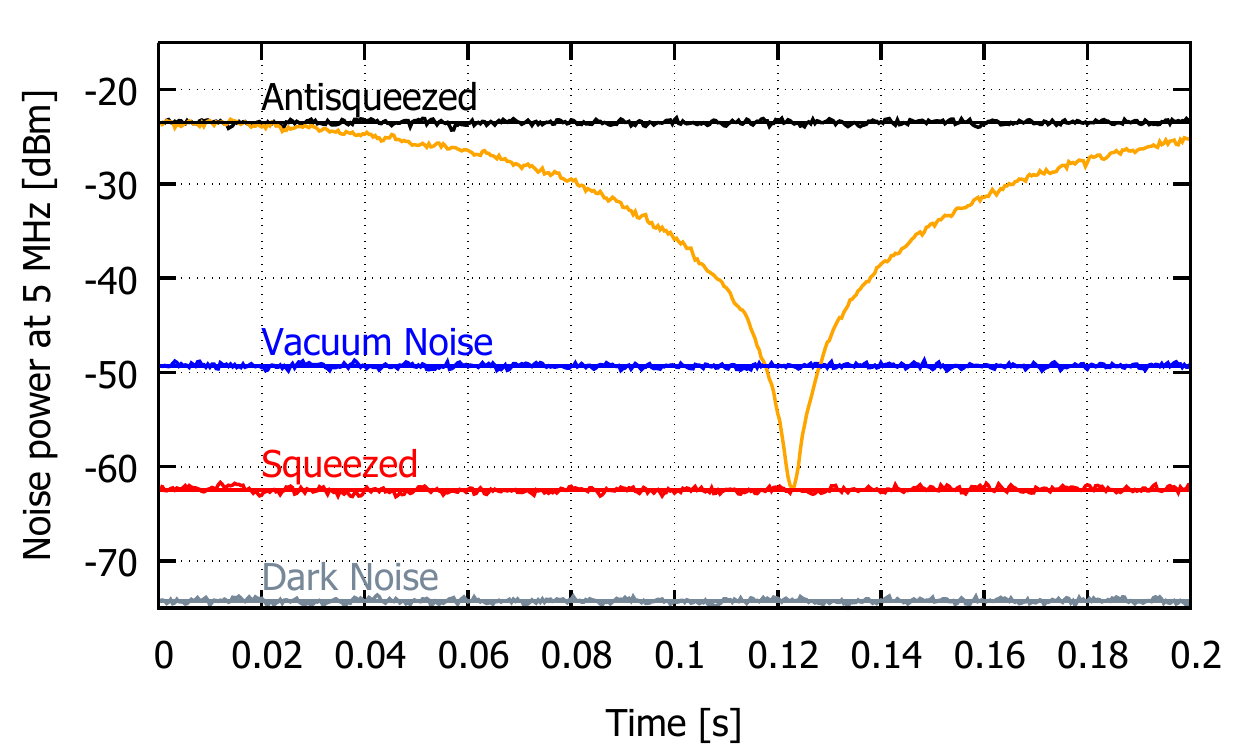}} 
\caption{Zero span measurements at Fourier frequency $f = 5\,\mathrm{MHz}$. With a pump power of $P_{\mathrm{pump}} = (12\pm1)\,\mathrm{mW}$ at $775\,\mathrm{nm}$, a squeeze factor of $13.1\pm0.05\,\mathrm{dB}$ below shot noise and an anti-squeeze factor of $25.8\pm0.05\,\mathrm{dB}$ was observed. (The error bars of these factors were due to imperfect stability of the local oscillator power during the measurement cycle.) The peaked curve was recorded with a slowly shifted local oscillator phase. 
The BHD dark noise was 24.9\,dB below the shot (vacuum) noise of the local oscillator power of 12\,mW and was not subtracted from the data. The resolution bandwidth was $\Delta f = 300\,\mathrm{kHz}$ and the video bandwidth $300\,\mathrm{Hz}$. 
}
\label{ZeroSpan}
\end{figure}
\begin{figure}[htbp]
\centering
\mbox{\includegraphics[width=\linewidth]{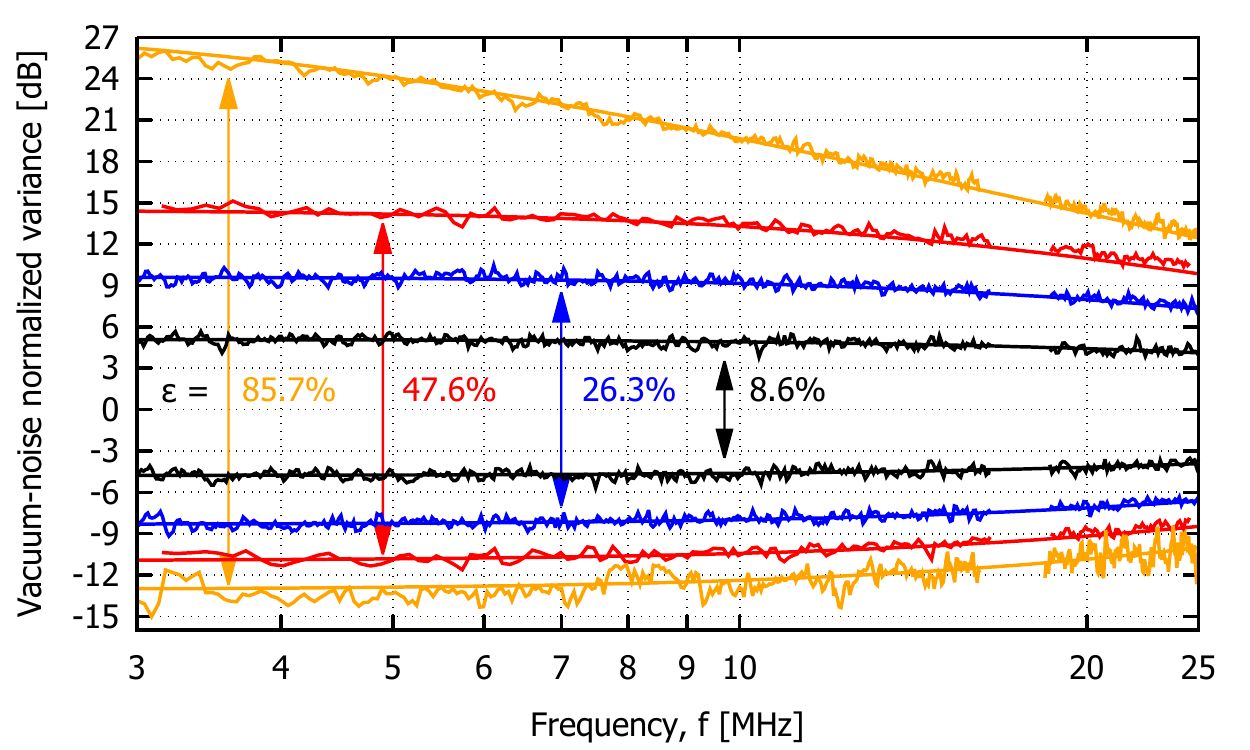}} 
\caption{Spectra illustrating the dependance of squeeze and anti-squeeze factors on the pump parameter. The pump powers were $85.7\,\%$, $47.6\,\%$, $26.3\,\%$ and $8.6\,\%$ of the oscillation threshold power, respectively. In this figure, the dark noise was subtracted from the data allowing for a straight forward comparison with our model. The fits were plotted for a PDC cavity FWHM of $109.8\,\mathrm{MHz}$ and an overall detection efficiency of $\eta = 0.952$. Phase noise was not included in the model. The resolution bandwidth was $\Delta f = 300\,\mathrm{kHz}$, and the video bandwidth $300\,\mathrm{Hz}$. All traces were recorded with a local oscillator power of 12\,mW, except for the one at $\epsilon = 85.7\,\%$. Here, the local oscillator power was reduced to 1\,mW to extend the linear response of the homodyne detector to large anti-squeezed noise powers (at the expense of a lower dark noise clearance).}
\label{Spectra}
\end{figure}

Figure \ref{ZeroSpan} shows zero span measurements at $f = 5\,\mathrm{MHz}$ taken with a resolution bandwidth of $\Delta f = 300\,\mathrm{kHz}$. The squeezing level is $13.1\pm0.05\,\mathrm{dB}$ below shot noise with an anti squeezing level of $25.8\pm0.05\,\mathrm{dB}$. The local oscillator power of $12\,\mathrm{mW}$ led to a dark-noise clearance of $24.9\pm0.05\,\mathrm{dB}$. The squeezing value was recorded with only $(12\pm1)\,\mathrm{mW}$ pump power at $775\,\mathrm{nm}$. The latter value is corrected for imperfect mode matching and corresponds to the external input power in the fundamental mode of the PDC cavity. The error is given by the uncertainty of the power-meter. For the peaked trace, we continuously scanned the phase of the local oscillator.\\
Figure \ref{Spectra} shows noise power spectra that were measured at different pump powers, normalized to the noise power of the local oscillator's shot noise. The data thus corresponds to quadrature amplitude variances of respective quantum uncertainties, where we denote the squeezed normalized variances as $\Delta^2 \hat X_{f \pm \Delta f /2}$ and the anti-squeezed normalized variances as $\Delta^2 \hat Y_{f \pm \Delta f /2}$. The frequency-span covers the range from $3\,\mathrm{MHz}$ to $25\,\mathrm{MHz}$. The smooth solid lines represent the following theoretical model, which holds for a lossy PDC cavity without phase noise operated below threshold \cite{Polzik1992,Aoki2006} assuming $f\! <\!<\! \Delta f$:
\begin{equation}
\Delta^2 \hat{X}_{f \pm \Delta f /2}= 1 - \eta \frac{4\sqrt{\epsilon}}{\left(1 + \sqrt{\epsilon}\right)^2+4\left( 2\pi f / \kappa \right)^2}\, ,
\label{Xspec}
\end{equation}
\begin{equation}
\Delta^2 \hat{Y}_{f \pm \Delta f /2}= 1 + \eta \frac{4\sqrt{\epsilon}}{\left(1 - \sqrt{\epsilon}\right)^2+4\left( 2\pi f / \kappa \right)^2}\,
\label{Yspec}
\end{equation}
where $\eta$ represents the overall detection efficiency of the initially pure quantum states;  $\epsilon= P_\mathrm{pump} / P_\mathrm{thrs}$ is the pump parameter with pump power $P_{\mathrm{pump}}$ and oscillation threshold power $P_{\mathrm{thrs}}$, and $\kappa$ is the decay rate of the PDC cavity at 1550\,nm.\\   
Fitting the model in Eqs.\,(\ref{Xspec}) and (\ref{Yspec}) to the eight spectra in Fig.\,\ref{Spectra} resulted in a full-width-half-maximum linewidth of $\kappa / (2\pi) = 109.8\,\mathrm{MHz}$, a total efficiency of $\eta=0.952$, and pump parameters of $85.7\,\%$, $47.6\,\%$, $26.3\,\%$ and $8.6\,\%$. The data around 17.5\,MHz were not included for fitting and omitted in the plot since they are overlapped with the phase modulation peak for cavity locking. Since the model represents our data quite well, we conclude that phase noise did not limit our measurements.

The strength of our squeezing was limited by $4.8\,\%$ optical loss. The imperfect mode matching at the BBS contributed most. We achieved a fringe visibility of $99\,\%$, resulting in $2\,\%$ loss. The remaining loss was due to imperfect photo diode quantum efficiency, imperfect propagation efficiency, and imperfect cavity escape efficiency. 
According to manufacturer specifications, the InGaAs photo diode quantum efficiency was $\eta_{pd} \approx 0.99$ (\emph{Laser Components}), the propagation efficiency was mainly limited by the transmission through altogether seven anti-reflection coated surfaces of three lenses and the PDC cavity coupler, yielding $\eta_{pr} \approx 0.99$, and the PDC cavity escape efficiency could be estimated to $\eta_{\mathrm{esc}} = T_1/ (T_1 + L) \approx 15\% / (15\% + 0.1\%) \approx 0.99$ (\emph{Laseroptik GmbH}). Here $T_1$ is the transmission of the PDC cavity coupler at 1550\,nm and $L$ is the cavity round trip loss at this wavelength, which includes two reflections from the anti-reflection coated surface of the PPKTP crystal, one transmission through the high-reflectivity coated back of the crystal and a small contribution from crystal absorption \cite{Steinlechner2013}.\\

The larger the PDC cavity built-up factor of the pump field, the lower is the required external pump power. The PDC cavity built-up factor is given by \cite{Hodgson2005,Thuering2009}
\begin{equation}
\frac{P_{\mathrm{cav}}}{P_{\mathrm{Pump}}} = \frac{\left( 1- R_{\mathrm{1}}\right)}{\left( 1-\sqrt{R_{\mathrm{1}} R_{\mathrm{2}}} V \right)^2}\,.
\label{PowerBuiltup}
\end{equation}
Here, $P_{\mathrm{cav}}$ and $P_{\mathrm{Pump}}$ are the intra-cavity and the external powers of the $775\,\mathrm{nm}$ pump field. $R_{\mathrm{1}}$ is the respective power reflectivity of the input mirror, and $R_{\mathrm{2}}$ the respective power reflectivity of the second cavity mirror. $V$ represents the propagation efficiency per half roundtrip at 775\,nm incorporating all losses from absorption and scattering. 
In our setup, $R_1 = (97.5 \pm 0.3)\%$, $R_2 = (99.955 \pm 0.004)\%$, and $V = V_{\rm AR} \cdot V_{\rm KTP} = (99.935 \pm 0.05)\%$, with a transmission through the AR coating of $V_{\rm AR} = (99.95 \pm 0.05)\%$, and a transmission through the PPKTP crystal of $V_{\rm KTP} = (99.985 \pm 0.005)\%$ \cite{Steinlechner2013}, which also takes into account scattering and absorption. 
These values we deduced from the measurement protocol of the coating company (\textit{Laseroptik GmbH}) and our own independent measurements. 
According to Eq.\,(\ref{PowerBuiltup}) they result in an intra-cavity power built-up factor of about 140. 
Increasing the in-coupling reflectivity to $R_{\mathrm{1}} = 99.8\,\%$ would result in a built-up factor of about $570$. Thus, the external pump power can be reduced by a factor of $570/140\!\approx\!4$ to about 3\,mW simply by increasing the reflectivity of the PDC coupler at 775\,nm.

\section{Discussion and Conclusion}

We demonstrated that the simultaneous conditions of double-resonance and quasi-phase-matching of a PPKTP squeezing resonator can be improved by coarse adjustment of the cavity length together with temperature fine-tuning.\\ 
For a given squeezing resonator - with given waist size and given optical loss - the external pump power required for achieving a certain squeeze factor, can be minimized by selecting the input coupler reflectivity close to impedance matching of the pump light. Our experimental setup was not optimized in such a way but still reached oscillation threshold at moderately low pump power of 14\,mW. Increasing the coupler reflectivity to $R_1{\rm(775nm)} = 99.8\,\%$ will increase the pump power built-up from 140 to 570: an external pump power of just 3\,mW can then produce the same spectrum with a squeeze factor of up to 13\,dB.\\
Further increasing the built-up factor at 775\,nm is straight forward. The AR- and the HR coating of the crystal in our setup were also not optimized for low external pump power. For instance, an improved AR coating that enables $V_{\rm AR} = 99.98\%$, an improved HR coating with $R_2 = 99.98\%$, and a coupler reflectivity of $R_1 = 99.9\,\%$ lead to $P_{\mathrm{cav}} / P_{\mathrm{Pump}} =  1100$, and the required external pump power reduces to 1.5\,mW, again without affecting the squeeze spectrum.\\ 
At a level of just a few milliwatts of pump power the overall requirement of optical power is already dominated by the balanced homodyne detector.
Some minimum optical power at the fundamental wave length is required for \emph{detecting} the squeezed uncertainty of the quadrature amplitudes \cite{Schnabel2017}. If the power is too low, the dark noise of the homodyne detector electronics limits the observable (useful) squeeze factor. Our experiment used up to 12\,mW for balanced homodyne detection. Since highly efficient second-harmonic generation from 1550\,nm to 775\,nm has been demonstrated \cite{Ast2011} we conclude that the light power consumption for generation and efficient detection of squeezed quadrature uncertainties can be below 15\,mW and clearly dominated by the detection scheme. To further reduce the required power, detection electronics with less dark noise need to be developed.

\section{Funding Information}

This work was partly supported by the European Research Council (ERC) project `MassQ' (Grant No. 339897).

\bigskip
\bibliography{sample}
\end{document}